**Erbium implanted silicon for solid-state quantum technologies**

*Mark A. Hughes\*, Naitik A. Panjwani, Matias Urdampilleta, Nafsika Theodoropoulou, Ilana Wisby, Kevin P. Homewood, Ben Murdin, Tobias Lindström and J. David Carey*

Dr M. A. Hughes, N. Theodoropoulou

Joule Physics Laboratory, School of Computing Science and Engineering, University of Salford, M5 4WT, UK

Email: *m.a.hughes@salford.ac.uk*

Dr N. A. Panjwani, Dr M. Urdampilleta

University College London, London Centre for Nanotechnology, Gower Place, WC1E 6BT, London, UK

Dr N. A. Panjwani

Berlin Joint EPR Lab, Fachbereich Physik, Freie Universität Berlin, D-14195 Berlin, Germany

Dr M. Urdampilleta

Institut Néel-CNRS-UJF-INPG, UPR2940 25 rue des Martyrs BP 166, 38042 Grenoble cedex 9, France

Dr I. Wisby, Dr. T. Lindström

National Physical Laboratory, Hampton Road, Teddington TW11 0LW, UK

Dr I. Wisby

Royal Holloway, University of London, Egham TW20 0EX, UK

Dr I. Wisby

Oxford Quantum Circuits Ltd. King Charles House 2nd Floor, Park End Street, Oxford, Oxfordshire, OX1 1JD, UK

Prof. Kevin P. Homewood, Prof. B. Murdin, Dr J. David Carey

Advanced Technology Institute, Faculty of Engineering and Physical Sciences, University of Surrey, Guildford, GU2 7XH, UK

Prof. Kevin P. Homewood

School of Materials Science and Engineering, Hubei University, Wuhan, 430062, Peoples Republic of China

Dr J. David Carey

Department of Electrical and Electronic Engineering, University of Surrey, Guildford, GU2 7XH, UK

**Abstract**

Erbium implanted silicon as a quantum technology platform has both telecommunications and integrated circuit (IC) processing compatibility. The electron spin coherence time of Er implanted Si with an Er concentration of $3\times10^{17}$ cm$^{-3}$ is measured to be ~10 µs at 5 K and the spin echo decay profile displays strong modulation due to super-hyperfine interaction with $^{29}$Si nuclei. Three independent measurements: temperature quenching of photoluminescence (PL), PL lifetime and photo-illuminated electron spin resonance (ESR) all indicate the presence of a previously unreported Er related defect state which can facilitate non-radiative relaxation from the Er exited state. This gives an energy level scheme analogous to that of the diamond NV centre, and implies that optical spin polarisation of the Zeeman ground state and high temperature operation of Er qubits in Er implanted Si may be feasible. The collective coupling strength between a superconducting NbN lumped-element microresonator and Er implanted Si with an Er concentration of $10^{17}$ cm$^{-3}$ at 20 mK was ~ 1 MHz.

# 1. Introduction

The optical fibre telecommunications network makes telecoms wavelength photons at 1.5 μm by far the best candidate for transferring quantum information over distance. $Er^{3+}$ intra 4f-shell transitions can be optically addressed at telecoms wavelengths which would allow transfer of quantum information over distance. The use of rare earth (RE) ions is well suited to overcome a paradox of quantum technology (QT) platform requirements: sufficient decoupling from the environment to avoid decoherence, but a strong enough interaction with the environment to allow addressing, readout and gating. The advantage of RE ions arises as they possess a partially filled 4f shell which is shielded from the environment by the outer 5s and 5p shells leading to extraordinary coherence times of 6 hours[1] and 4.4 ms[2] for optically detected nuclear spin and electron dipole transitions, respectively; however, even with their atomic scale shielding, long lived entanglement between RE dopants in a solid matrix has been observed,[3] and entanglement between internal degrees of freedom of single RE ions can still exist up to thousands of K, making this one of the most stable known entanglements.[4]

A practical quantum computing architecture developed in Si will move from a one-off device to production far quicker than for any other candidate, and features can be patterned in Si on the scale required for many quantum device architectures. Ion implantation of Si is a well understood technology in IC fabrication, and commercial adoption of new technologies tend to favour those based on established fabrication platforms and techniques. Recently, increases in coherence times by several orders of magnitude have been demonstrated in donor impurities in silicon by using isotopically pure $^{28}$Si.[5] However, these donor impurities do not interact with light at telecommunications wavelengths, which is critical for many quantum communication schemes. Given expected improvements in $T_2$ by using $^{28}$Si, optimising processing for the appropriate Er-related centre[6] and reducing Er concentration, Er implanted Si is potentially the only know QT platform with telecoms addressability, long $T_2$ and IC tooling compatibility. The spin state of a single Er ion implanted into a silicon single electron transistor has been optically addressed and electrically readout,[7] whereas, the spin state of a single Er ion in $Y_2SiO_5$, coupled to a silicon nanophotonic cavity, can be readout optically with a single shot.[8] This demonstrates that Er implantation is compatible with a potential quantum computing architecture and that Er could be integrated in to quantum communication and information processing schemes.

Arguably the most sophisticated quantum computers to-date use superconducting (SC) circuits. Using Er spin ensembles can also provide a telecommunications interface for SC resonator qubits,[9] and a single ensemble could be used to store many qubits by using holographic encoding.[10]

The NV centre in diamond is perhaps unique among solid state defects in that it can be optically spin polarized to give an effective spin temperature of ~ 0 K and coherence times ($T_2$) of a few ms, while the surrounding diamond lattice is at room temperature.[11] All the other viable solid-state quantum technology platforms must be cooled to ~ 0 K to have adequate coherence times. However, the lack of IC processing capability in diamond makes it far less viable for large scale quantum computers, and its visible wavelength operation makes it less viable for quantum communication schemes.

Here we report the first coherence measurements of Er implanted Si, or, in fact, any implanted Er, in the form of spin echo measurements, we demonstrate the first of coupling between Er implanted in Si and a superconducting resonator. We also report a previously unreported Er related defect state which leads to an energy level scheme analogous to that of the diamond NV centre, and gives a mechanism for optical spin polarisation of the Zeeman ground state which could lower the effective spin temperature and increase the working temperature at which reasonably long coherence times can be obtained.

## 2. Experimental

### 2.1 Sample preparation

Sample 1, with $10^{19}$ cm$^{-3}$ Er and $10^{20}$ cm$^{-3}$ O, was prepared by implanting Er with a total areal dose of $2.6\times10^{15}$ cm$^{-2}$ and O with a total areal dose of $1.3\times10^{16}$ cm$^{-2}$ at 77 K into one face of a <100> oriented 8000 ± 500 Ωcm Si wafer of 500 µm thickness supplied by Topsil. Sample 1 was annealed with recipe *a,* which consisted of a 450°C for 30 min anneal to smooth the crystalline-amorphous interface, a 620°C for 180 min anneal to recrystallize the amorphized region and a 850°C for 30 s anneal to activate the Er. It was previously found that annealing at 850°C significantly increased the ESR signal strength.[12] Sample 2, with $3\times10^{17}$ cm$^{-3}$ Er and $10^{20}$ cm$^{-3}$ O, and sample 3, with $10^{17}$ cm$^{-3}$ Er and $10^{20}$ cm$^{-3}$ O, were prepared by implanting at room temperature into both faces of a <100> oriented 5000-9999 Ωcm Si wafer of 50 µm thickness supplied by Si-Mat. Samples 2 and 3 were annealed with recipe *b* (750°C for 2 min) and recipe *c* (600°C for 167 min, then 850°C for 30 s), respectively. For all samples, O and Er ions were implanted at a range of energies to give a flat concentration profile down to a depth of around 1.5 µm, see supplementary **Figure S1**. Isotope specific implantation was used so that only the zero nuclear spin $^{166}$Er was implanted.

### 2.2 ESR and spin echo measurements

CW and pulsed ESR measurements were performed in a Bruker E580 ESR spectrometer. All ESR measurements were recorded with the magnetic field, $B_0$, parallel to the [001] direction of the wafer with an uncertainty of ±5°. When using Er concentrations $\leq 3\times10^{17}$ cm$^{-3}$ and recipe *a*, no spin echo signal could be observed. We switched to 50 µm thick wafers in a effort to maximise the ESR resonator's Q factor. We also tried two different recrystallization strategies: shorter time, higher temperature and longer time, lower temperature. These were 750°C for 2 min (recipe *b*) and 550°C for 335 min (recipe *d*), with no activation anneal; recipe *d* resulted in no measurable ESR resonances. However, recipe *b* resulted in strong, narrow ESR resonance lines, and a measurable echo signal. With samples 1 and 2, the Q factor of the ESR resonator was ~600 and ~9000, respectively, due to metallic doping at the high Er concentrations in sample 1 and from higher sample volume. For pulsed measurements the Q factor was detuned to ~100 for both samples.

## 2.3 Optically Modulated Magnetic Resonance (OMMR)

For OMMR measurements we used a Bruker EMX ESR spectrometer, incorporating a super high-Q resonator with optical access. A 20 mW external cavity laser tuneable from ~1490 to 1620 nm, was fibre coupled to a C-band erbium doped fibre amplifier (EDFA) with a gain bandwidth of ~1510-1600 nm and an output of up to ~150 mW. The output of the EDFA was sent through a linear polarizer, parallel to the $[1\bar{1}0]$ direction of the Er implanted Si wafer, then modulated at ~30 Hz with a mechanical chopper. The output signal from the ESR spectrometer's on-board lock-in amplifier was used as the input to an external lock-in amplifier (SRS830) and referenced to the mechanical chopper. OMMR spectra were produced by sweeping the external cavity laser wavelength and reading the SRS830 lock-in signal to give a spectrum of the ESR signal that has been modulated by the laser. A microwave power of 2.1 mW, a field modulation of 100 kHz, a time constant of 5 ms and a microwave frequency of 9.37 GHz was used for all OMMR measurements.

## 2.4 PL measurements

PL spectra were obtained by cooling the sample in a cold finger $LH_2$ cryostat at 15 K. Excitation was by a 462 nm 50 mW laser diode and the generated fluorescence was dispersed in a Bentham TMc300 monochromator and detected with an IR PMT coupled with standard phase sensitive detection. All spectra were corrected for the system response.

**2.5 Superconducting resonator coupling**

Superconducting resonator coupling measurements were performed in a dilution refrigerator, fitted with a vector magnet, at 20 mK. A superconducting lumped element micro-resonator was fabricated by sputtering 200 nm of NbN, patterned by standard e-beam lithography, onto an R-cut sapphire substrate. Sample 3 was placed face down on the microresonator and the microresonator was placed in a magnetic field that was stepped from zero to 93 mT. The power in the resonator was ~3 pW. At each magnetic field the microwave transmission coefficient, $S_{21}$, was measured using a vector network analyser (VNA). This was repeated for magnetic field orientations between 0° and 160° in steps of 5°, with 0° corresponding to $B_0$ parallel to the face of the resonator and sample. The magnetic field was rotated around the [110] crystal axis of the sample. Numerical fitting of the $S_{21}$ response of the microresonator was used to extract the total measured loss tangent $\tan \delta_{tot} = 1/Q_{tot}$, where $Q_{tot}$ is the total measured Q factor and $\tan \delta_{tot} = \tan \delta_c + \tan \delta_{diel} + \tan \delta_B + \tan \delta_{ions}$, are the loss tangents due to coupling to the transmission line, dielectric losses, the external magnetic field and the Er ions, respectively. Numerical fitting was then used to extract $\tan \delta_{ions}$.

**3. Results and discussion**

**3.1 Spin Echo Measurements**

We used three different annealing recipes- *a*, *b* and *c* and isotopically selective implantation of $^{166}$Er to fabricate samples with Er concentrations between $10^{17}$ and $10^{19}$ cm$^{-3}$ for electron spin resonance (ESR), spin echo and SC resonator coupling measurements. When implanted into Si, Er exists in its usual 3+ oxidation state.[13] Oxygen was co-implanted to a concentration of $10^{20}$ cm$^{-3}$ for all samples and is required to generate narrow Er-related ESR[14] and photoluminescence (PL)[15] lines by the creation of various O coordinated Er (Er-O) centres. Previous measurements of the angular dependence of the Er-related ESR lines in Er implanted Si have identified a number of different Er-O ESR centres: three monoclinic centres labelled OEr-1, OEr-1' and OEr-3, and three trigonal centres labelled OEr-2, OEr-2'and OEr-4.[14, 16, 17] None of the ESR centres are optically active.[12] Also, Zeeman measurements of molecular-beam epitaxy (MBE) grown Er doped Si have identified an orthorhombic Er-O centre,[18] which is not ESR active.[12] The *g* tensors of these centres are given in Table 1.

**Table 1** The g tensors of Er doped Si determined by ESR and Zeeman measurements. The ESR centres OEr-1, OEr-2 etc are referred to as 1,2 etc. in subsequent ESR spectra.

| Centre | Symmetry | $g_x$ | $g_y$ | $g_z$ | Ref. |
|---|---|---|---|---|---|
| ESR | | | | | |
| OEr-1 | Monoclinic $C_{1h}$ | 0.8 | 5.45 | 12.6 | [14] |
| OEr-1' | Monoclinic $C_{1h}$ | 0.8 | 5.45 | 12.55 | [14] |
| OEr-3 | Monoclinic $C_{1h}$ | 1.09 | 5.05 | 12.78 | [14] |
| OEr-4 | Trigonal $C_{3v}$ | 2.0 | 6.23 | 6.23 | [14] |
| OEr-2 | Trigonal $C_{3v}$ | 0.45 | 3.46 | 3.22 | [14] |
| OEr-2' | Trigonal $C_{3v}$ | 0.69 | 3.24 | 3.24 | [14] |
| Zeeman | | | | | |
| Er-1 | Orthorhombic $C_{2v}$ | 0 | 0 | 18.4 | [18] |

By comparison to these angular dependencies we can attribute the CW ESR line at 963 G from the sample with $10^{19}$ cm$^{-3}$ Er and $10^{20}$ cm$^{-3}$ O with annealing recipe *a* (sample 1), shown in **Figure 1a**, to the OEr-3 monoclinic centre. These Er-related ESR lines have been attributed to an Er$^{3+}$ centre based on similar *g* tensors to Er doped Y$_2$O$_3$, which has the same crystal structure as Er$_2$O$_3$, and EXAFS measurements of Er and O implanted Si which found a similar Er-O bond length to that in Er$_2$O$_3$.[14, 17] **Figure 1a** also shows that the spin echo peaks at the same B$_0$ as the Er-related ESR resonance at 963 G, which has a full width at half maximum (FWHM) of 7 G. Note that a small angular deviation in the [1$\bar{1}$0] axis can cause significant shifting of the ESR resonances. A spin echo signal was also present off-resonance, and was also present in empty tube measurements, we therefore treated the off-resonance echo as a background and subtracted it, as illustrated in supplementary **Figure S2**. **Figure 1b** shows the integrated intensity of the on-resonance echo peak as a function of the delay time between the π/2 and π pulses (t$_{12}$). Fitting to an exponential decay gives an estimated T$_2$ of 180 ± 80 for sample 1.

**Figure 1d** shows the CW ESR and field dependent echo for the sample with 3×10$^{17}$ cm$^{-3}$ Er and 10$^{20}$ cm$^{-3}$ O with annealing recipe *b* (sample 2). There are two main resonances at 867 and 934 G, respectively, with weaker resonances at 892 and 964 G. Using the well described angular dependence of Er implanted Si ESR,[14, 16, 17] these resonances are also attributed to the OEr-3 and OEr-1 centres, as shown; all of these resonances are visible in the echo spectrum. **Figure 1d** shows the echo intensity as a function of B$_0$ for various t$_{12}$ between 0.14 and 2.24 μs for sample 2. The on-resonance echo signal disappears below the detection limit then reappears with increasing t$_{12}$, indicating the presence of very strong electron spin echo envelope modulation (ESEEM).

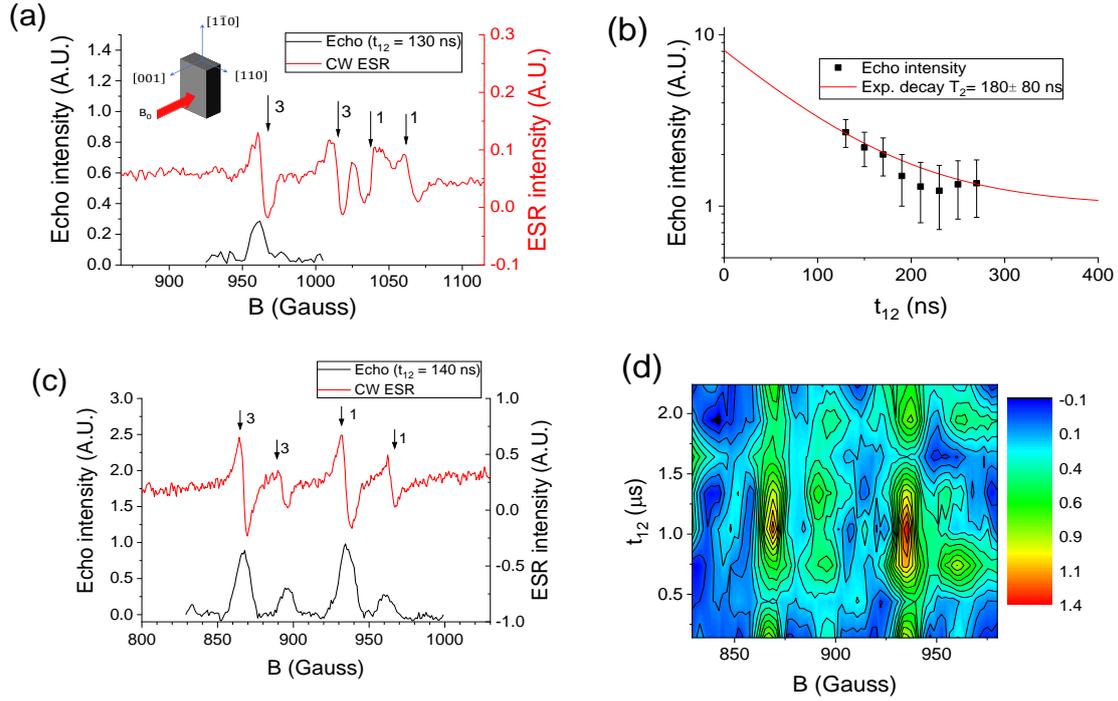

**Figure 1** a) CW ESR and field dependent echo signals for $10^{19}$ cm$^{-3}$ Er and $10^{20}$ cm$^{-3}$ O with annealing recipe *a* (sample 1); limitations of the measurement system meant that delays shorter than 130 ns could not be used. The inset illustrates the alignment of the sample 1 and 2 during ESR measurements. b), Integrated intensity of the echo signal as a function of delay time for sample 1. c), CW ESR and field dependent echo signal for $3\times10^{17}$ cm$^{-3}$ Er and $10^{20}$ cm$^{-3}$ O with annealing recipe *b* (sample 2). d), Contour plot showing the echo intensity as a function of magnetic field at various $t_{12}$ for sample 2. All CW ESR measurements were made at 10 K, all echo measurements at 5 K, and the microwave frequency was 9.61 GHz. The ESR resonances are attributed to the previously identified ESR centres in Table 1.

In **Figure 2a** we show the echo decay profile, at fixed $B_0$, for sample 2. The background echo signal was subtracted as shown in supplementary **Figure S3** to give an on-off resonance echo decay. As can be seen from **Figure 1c**, a subtraction of the echo decay at 850 G from that at 867 G, should leave only the echo component attributable to implanted Er. The echo decay displays strong superimposed oscillations from the ESEEM effect[19] caused by superhyperfine coupling with neighbouring nuclear spins; similar oscillations were observed in Er:CaWO$_4$, but were significantly weaker than those seen here.[20] Since the 4f wavefunction is highly localised, the superhyperfine coupling between a RE and a neighbouring nuclear spin is usually regarded as magnetic dipole-dipole only.[21] The echo decay of an isolated $Er^{3+}$ ion (effective electron spin S = ½) in proximity to a $^{29}Si$ nuclei (nuclear spin I = ½) can be described as follows,[19]

$$I(t_{12}) = I_0 e^{-\left(\frac{2t_{12}}{T_2}\right)^x}\left[1 - 2k\sin^2(2\pi\nu_\alpha t_{12}/2)\sin^2(2\pi\nu_\beta t_{12}/2)\right]. \qquad (1)$$

The first term is the empirical Mims equation,[22], which describes the echo decay in the absence of nuclear coupling where $T_2$ is the spin coherence time, and *x* is an exponential stretch factor which is determined by spin dynamics. The term in square brackets describes the ESSEM modulation, where *k* is the modulation index, $\nu_\alpha$ and $\nu_\beta$ are the $^{29}Si$ nuclear resonance frequencies for the two possible $Er^{3+}$ electron spin orientations (S = ±½). Equation (1) encompasses weaker sum ($\nu_+ = \nu_\alpha + \nu_\beta$) and difference ($\nu_- = \nu_\alpha - \nu_\beta$) frequency components. In **Figure 2a** fitting to Equation (1) yields $\nu_\alpha$ of 0.70, which is close to the Larmor frequency ($\nu_L$) of 0.76 MHz for $^{29}Si$ at this magnetic field, and $\nu_\beta$ of 1.63 MHz; these frequencies can be seen in the FFT in **Figure 2b**. The FFT peaks are broadened by the relatively short sampling time. The deviation of $\nu_\beta$ from $\nu_L$ indicates stronger super-hyperfine coupling between Er electron spins and $^{29}Si$ nuclear spins. The FFT of the fitted decay show the $\nu_+$ and $\nu_-$ components; these cannot be resolved on the FFT of the measured decay because of insufficient signal-to-noise.

The modulation of the on-off resonance decay is very strong with fitting yielding *k* = 0.45. The fit also yielded a $T_2$ of 9 ± 3 µs. Our value of $T_2$ of ~10 µs at 5 K compares to ~5 µs at 5 K (~50 µs at 2.5 K) for ~$10^{16}$ $cm^{-3}$ Er doped $CaWO_4$.[20] and 1.6 µs at 1.9 K for ~$2\times10^{18}$ $cm^{-3}$ Er doped $Y_2SiO_5$.[23] This is a promising comparison given the difficulty of removing defects after implantation that could lead to decoherence. Further optimisation of the recrystallization process, reductions in Er concentration and isotopic purification of the Si may lead to coherence times applicable to quantum communication and computation.

The saturation recovery profile is shown in **Figure 2c** gives a spin relaxation time, $T_1$, at 5 K, ~1 ms; the background subtraction is shown in supplementary **Figure S4**. Given that $T_1 \gg T_2$, it is concluded that $T_2$ is limited by local field fluctuations.[22] For both donors in silicon and RE doped transparent crystals, the ESEEM effect is thought to be caused largely by nuclear spins in very close vicinity to the echo producing centre. For example in both P[24, 25] and Bi[26] doped Si, the ESEEM effect is attributed to the four nearest Si lattice positions, in $Ce^{3+}$ doped $CaWO_4$, the ESEEM of $Ce^{3+}$ can be accurately modelled using the closest ten lattice positions of surrounding W atoms.[19] Spectral diffusion can be caused by various electron[27] and nuclear[28] spin flip-flop process; in Er implanted Si, nuclear induced spectral diffusion is most likely since $^{29}Si$ nuclear spins are present and for the electron spins $T_1 \gg T_2$. The nuclear spins involved in ESEEM, which experience large hyperfine fields, cannot flip flop their spins and contribute to spectral diffusion due to conservation of energy, whereas the nuclear spins involved in spectral diffusion must experience very weak hyperfine fields to allow flip-flops, and therefore consist of a separate, larger, group of nuclei that are further from the echo producing centre.[24] The observation that fitting yielded $x \sim 1$ in **Figure 2a**, is somewhat unexpected and indicates no significant spectral diffusion occurs.[2] Due to our isotope specific implantation, the only nuclear spins in our sample are from $^{29}Si$.

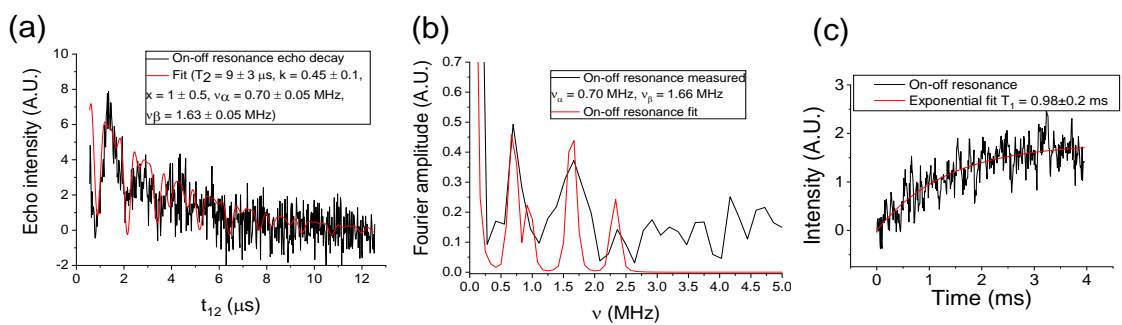

**Figure 2** a) Spin echo decay profile at a $B_0$ of 867 G fitted to Equation (1). b) FFT of measured and fitted decay profile. c) saturation recovery at 867 G fitted with a single exponential fit to give $T_1$ of ~1 ms. All measurements were on sample 2 ($3\times10^{17}$ cm$^{-3}$ Er and $10^{20}$ cm$^{-3}$ O, with annealing recipe *b*) at 5 K, and the microwave frequency was 9.61 GHz.

**3.2 Photoluminescence**

**Figure 3a** shows the PL spectrum from sample 1 at 15 K. PL peaks from this sample that we have previously demonstrated originate from the orthorhombic centre[12] are indicated. **Figure 3b** shows the Arrhenius plot of the PL intensity of the main peak at 6505 cm$^{-1}$. The temperature quenching from Er implanted Si typically displays two activation energies, with slow quenching of the PL at temperatures up to ~150 K and rapid quenching after ~150 K, due to Auger and back-transfer quenching mechanisms, respectively, which are explained elsewhere,[29] and can be described by Equation (2), below, but without the $A_1 e^{(-E_{NT}/k_B T)}$ term. However, **Figure 3b** shows an increase in PL intensity as temperature is increased from ~150 K. This negative thermal quenching has not been reported before in Er implanted Si, to the best of our knowledge, and implies population of the initial state, in this case the Er$^{3+}$ $^4I_{13/2}$ state, by thermal excitation from a middle state, which we will refer to as the Er-O deep state, between the initial and final states, and is described by Equation (2) for the case of one middle state.[30] There is an excellent fit to Equation (2) which gives Auger quenching ($E_{AQ}$), back transfer ($E_{BT}$) and negative thermal quenching ($E_{NT}$) activation energies of 900 (110), 2600 (320) and 1520 (190) cm$^{-1}$ (meV), respectively. This compares to previous reports of $E_{AQ}$ and $E_{BT}$ of 160 (20) and 1210 (150) cm$^{-1}$ (meV), respectively.[13] The PL lifetime of sample 1 was less than the system response of ~1 µs, whereas the PL lifetime of Er implanted Si is usually ~1 ms.[31] This decrease in lifetime could be caused by clustering, but it is also consistent with the Er-O deep state indicated by Arrhenius plot in **Figure 3a** which could facilitate non-radiative decay.

$$I(T) = I_0 \frac{1+A_1 e^{(-E_{NT}/k_B T)}}{1+C_1 e^{(-E_{BT}/k_B T)}+C_2 e^{(-E_{AQ}/k_B T)}} \qquad (2)$$

PL from samples similar to our sample 1 have been previously reported,[14, 31] the only differences to our processing was that the Er implant chain went up to 5 MeV, rather than our 4 MeV, the final activation anneal was 900°C, rather than our 850°C and the P-doped FZ wafer had a resistivity of ~220 Ωcm, rather than our wafer with a resistivity of 8000 ± 500 Ωcm. The PL spectrum was similar to ours; however, the Arrhenius plot of the PL showed no negative thermal quenching and $E_{AQ}$ and $E_{BT}$ of 160 and 1210 cm$^{-1}$, respectively.[31] Time decay of the PL revealed fast (80 μs) and slow (800 μs) lifetime components. [14, 31] This very similar sample therefore has no evidence of an Er-O deep state, unlike our sample 1. The most likely reason is the significantly lower P concentration of our wafer since many Er-O defect states are known to form in Er implanted Si and their presence and energy depends strongly on the presence of other impurities, including P.[13] This higlights how sensitive the energy level structure and optical properties of the Er-O centres are to processing conditions. The fraction of implanted Er that becomes optically active varies between 1 and 10% for different studies;[13] while relatively low, this large range of yields is also consistent with a high sensitivity to processing conditions, which leaves a lot of scope for optimising the processing conditions. There are no estimates of the implantation yield for ESR active centres.

**Figure 3c** shows the PL mechanism for sample 1, which is based on the well-established PL mechanism for Er implanted Si,[13] where above bandgap energy excitation generates excitons which are trapped at Er coordinated O (Er-O) donor traps which introduces a state, $E_{DT}$, ~1210 cm$^{-1}$ below the conduction band, as measured by deep level transient spectroscopy (DLTS). This value of $E_{DT}$ and our $E_{BT}$ from fitting in **Figure 3b** indicates that the bottom of the Er$^{3+}$ $^4I_{15/2}$ manifold is offset by and energy ,$E_{off}$, of ~1500

(185) cm$^{-1}$ (meV) below the top of the valence band, whereas previously reported values of $E_{BT}$ usually put the $^4I_{15/2}$ manifold just above the VB.[13]

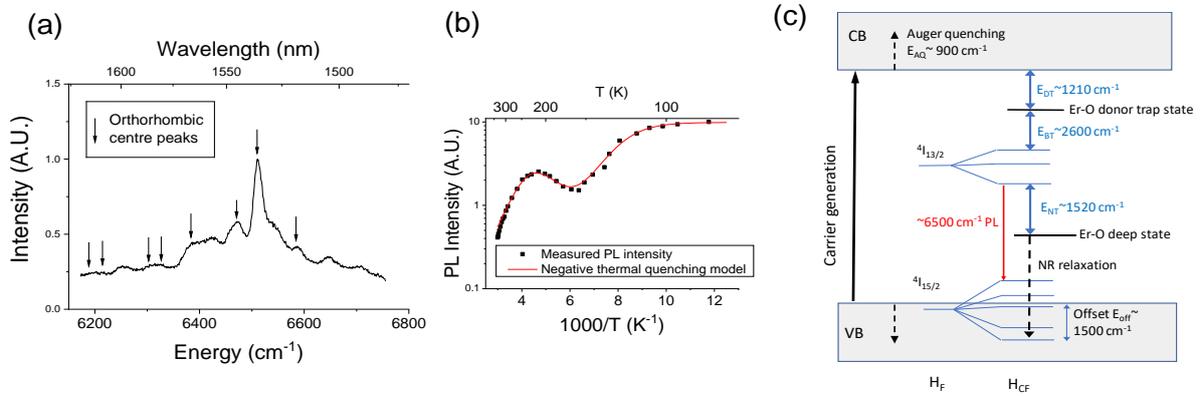

**Figure 3** a) Photoluminescence spectrum of $10^{19}$ cm$^{-3}$ Er and $10^{20}$ cm$^{-3}$ O with annealing recipe *a* (sample 1) at 15 K excited at 462 nm. b) Arrhenius plot of the PL intensity from sample 1 at various temperatures fitted to Equation (2). c) Proposed PL mechanism for sample 1.

### 3.3 Optically modulated magnetic resonance

**Figure 4a** shows a contour plot built from many OMMR spectra at various magnetic fields. The various OMMR spectra appear similar, with four broad bands at 6230, 6380, 6530 and 6610 cm$^{-1}$, which vary in intensity with the magnetic field. **Figure 4a** shows a broad optically generated ESR resonance appears which appears as two bands at 850 and 1170 G, both of which are positive since the second lock-in gives the absolute value of the modulated ESR signal. Because of the lock-in technique used in ESR spectrometers, ESR spectra have zero signal at the maximum of the magnetic centre's absorption; for the same reason, the OMMR spectra are minimised at ~ 960 G, which we refer to as the zero-crossing region. **Figure 4b** show a close-up of the zero-crossing region which shows how the four broad bands evident in **Figure 4a** narrow and split into narrower bands in the zero-crossing region. **Figure 4c** shows the OMMR spectra from various magnetic fields inside (951 and 969 G) and outside (1180 G) the zero-crossing region. It is clear how the narrow peaks inside the zero-crossing region align with broad peaks outside the zero-crossing region. The reason the OMMR peaks broaden outside the zero-crossing region is probably an artefact of the measurement system since the zero-crossing region is particular to the lock-in technique used in ESR spectrometers. Peaks that we previously demonstrated must originate from transitions to the $^4I_{13/2}$ excited state manifold of the orthorhombic Er-O centre[12] are highlighted. We previously[12] proposed a model for the OMMR mechanism involving transitions from the Si valence band (VB) to the $^4I_{13/2}$ manifold which relax back to the Zeeman ground state of the $^4I_{15/2}$ manifold and augment the ESR signal. The bottom of the $^4I_{15/2}$ manifold must be some offset energy, $E_{off}$, below the top of the VB because the highest energy PL transition, in **Figure 3a**, should have the same energy as the lowest energy OMMR transition (equivalent to absorption) in **Figure 4c**, yet there is a difference $E_{off}$ = 227 cm$^{-1}$. **Figure 4d** shows OMMR spectra at 10 K and 4 K, both at magnetic fields outside the zero-crossing region. The broad peaks can be aligned by adding 80 cm$^{-1}$ to the 10 K spectrum, showing that $E_{off}$ at 4 K is 197 cm$^{-1}$ (80 cm$^{-1}$ less than at 10 K), indicating that $E_{off}$ has a temperature dependence. An $E_{off}$ which increases with temperature is consistent with the model from PL temperature quenching in **Figure 3b** where $E_{off}$ ~1500 cm$^{-1}$ was derived from $E_{BT}$ which came from fitting the temperature quenching at >200 K. Peaks at 6285 and 6360 cm$^{-1}$ are yet to be identified but could originate from another Er-O center with a higher $E_{off}$ than the orthorhombic Er-O center. When the temperature was reduced from 10 K to 4 K the OMMR intensity increased and there were no narrow peaks visible at the zero crossing region, indicating that a low OMMR signal intensity is required to observe the narrow OMMR peaks

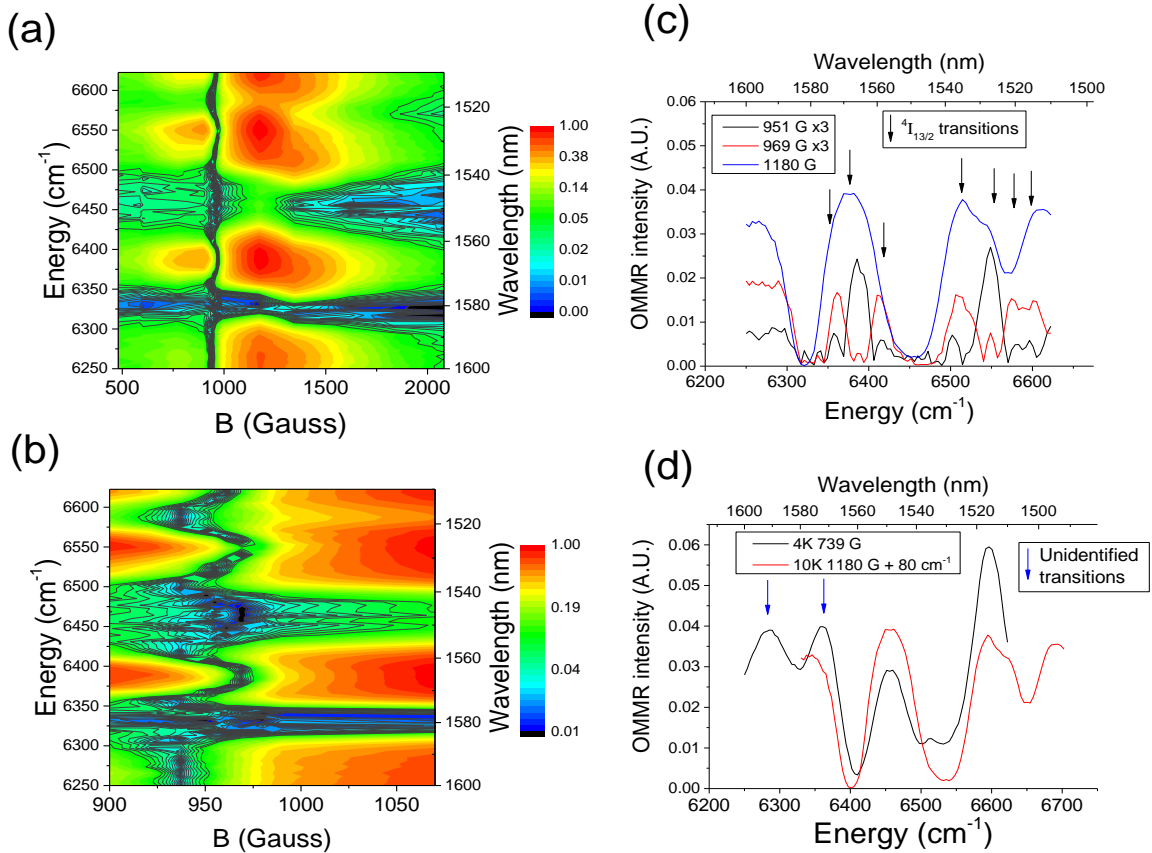

**Figure 4 a)** Contour plot created from various OMMR spectra at magnetic fields between 480 and 2300 G at 10 K. b) close up of the zero-crossing region at 10 K. c) OMMR spectra at various magnetic fields inside and outside the zero-crossing region, arrows indicate the peaks associated with transitions to the $^4I_{13/2}$ manifold, temperature was 10 K. d) OMMR spectra at 4 K and 10 K, shifted by 80 cm$^{-1}$ to illustrate the alignment of peaks, arrows indicate unidentified OMMR peaks. $B_0$ was approximately aligned with the [110] axis in panels.

**Figure 5a** shows the unilluminated ESR spectrum of sample 1 with the well-known narrow monoclinic and trigonal ESR lines identified. Under illumination at 1535 nm a strong broad ESR resonance with a width of ~150 G and an integrated intensity three orders of magnitude greater than the narrow unilluminated ESR resonances is visible.

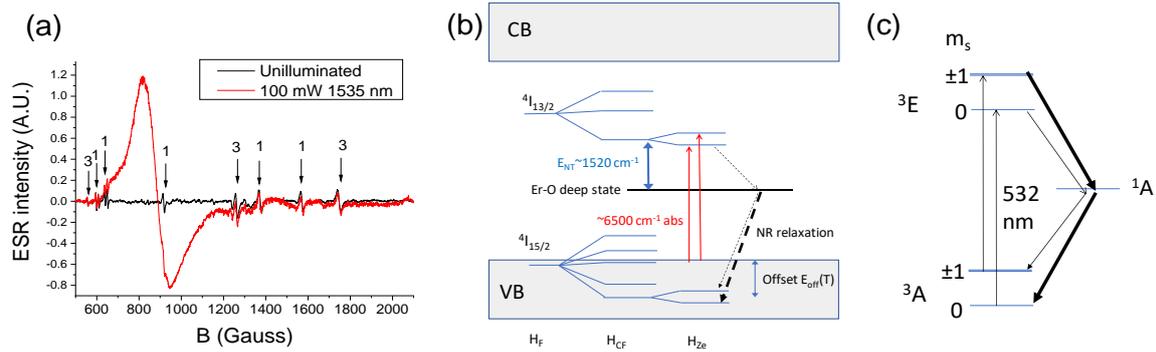

**Figure** 5a) Unilluminated and illuminated ESR spectrum from sample 1 with B approximately aligned with the [110] axis. b) Proposed mechanism for the OMMR/illuminated ESR signals from sample 1. c) Optical spin polarization mechanism of the NV centre in diamond.

The very high relative intensity of the illuminated ESR resonance implies a very efficient mechanism to populate the Zeeman ground state, but intra 4f-orbital relaxation rates are very low, particularly at 4 K, which excludes a direct $^4I_{13/2}$ to $^4I_{15/2}$ transition to repopulate the Zeeman ground state. The Er-O deep state suggested by PL temperature quenching and lifetime measurements can explain the very high relative intensity of the illuminated ESR resonance with the addition that the non-radiative relaxation is spin selective. This mechanism is illustrated in **Figure 5b** and is reminiscent of the optical spin polarization mechanism for NV centres shown in **Figure 5c.** The very high relative intensity of the illuminated ESR resonance implies that the mechanism could be as efficient as in NV centres, and could therefore lead to a very low effective spin temperature. Various schemes have been used previously to spin polarise REs to the Zeeman excited state, such as using stimulated emission in $Er^{3+}:Y_2SiO_5$,[32] circularly polarised light in $Ce^{3+}:YAG$,[33] and by tuning the branching ratio between Zeeman states in ground and excited state manifolds in $Nd^{3+}:YVO_4$.[34] However, these schemes don't have the same analogy to the NV centre mechanism that the Er-O mechanism in **Figure 5b** has.

## 3.4 Superconducting resonator coupling

The implanted face of sample 3 ($10^{17}$ Er cm$^{-3}$ and $10^{20}$ O cm$^{-3}$, annealing recipe *c*) was placed in contact with the superconducting NbN lumped-element microresonator on R-cut Al$_2$O$_3$ shown in **Figure 6a**, which had a centre frequency $\omega_r/2\pi = 3.04$ GHz, see experimental section. **Figure 6b** shows the loss tangents due to coupling to Er ions (tan $\delta_{ions}$) as function of $B_0$ and orientation. There is a single narrow resonance, with a FWHM of 50±10 G, that varies smoothly between 740 and 870 G depending on the $B_0$ orientation. There is also a very broad resonance centred at 500 G and at $B_0$ // [001] (0° orientation), the resonance shifts to 600 G at 50° $B_0$ orientation; we simulated the angular dependence of the six ESR centres (three trigonal, three monoclinic) previously identified Er and O implanted Si system[14], see Table 1, but found no correspondence with this broad resonance. The narrow resonance had a remarkable correspondence with the trigonal OEr-2' centre identified in ref. [14] with $g_\parallel = 0.69$ and $g_\perp = 3.24$, which is shown in the simulation in **Figure 6c**. The two other resonances are also visible in this $B_0$ range but are significantly weaker, which explains why only one resonance is observed in the microresonator measurement. A higher $B_0$ range shows the positions of all three expected ESR resonances with trigonal symmetry in the simulation in **Figure 6d**. Only one previously identified Er centre is evident in the hybrid measurements. This could be due to preferential coupling of the trigonal centre, or the trigonal centre has a shorter $T_1$ than the other centres at 20 mK, which prevented saturation.

The Q factor of a resonator coupled to an ensemble of spins can be modelled as a single mode harmonic oscillator according to

$$Q_{tot} = \frac{\Delta^2 + \gamma^2}{2g_{col}^2 \gamma + \kappa(\Delta^2 + \gamma^2)} \omega_r, \tag{3}$$

where $\Delta$ is the detuning from the spin resonance peak, $\gamma$ is the spin linewidth, $\kappa$ is the cavity linewidth = $2\pi\omega_r/Q_{tot}$ = 0.56 MHz for the 0° orientation and was independently measured away from the resonance for each $B_0$ orientation, $Q_{tot}$ is the total measured cavity Q, and $g_{col}$ is the collective coupling strength. Supplementary **Figure S5** shows the fitting of Equation (3) to the $Q_{tot}$ for the 0° orientation which gives $g_{col}/2\pi$ = 1 MHz and $\gamma/2\pi$ = 80 MHz. The average for all $B_0$ orientations was $g_{col}/2\pi$ = 1.1 ± 0.3 MHz and $\gamma/2\pi$ = 85 ± 25 MHz. The coupling strength of an individual spin to the SC resonator is given by $g_i = g_{col}/\sqrt{N}$, where N is the number of spins coupled to the resonator; using the number of Er ions above the microresonator (~$3.7\times10^{10}$) gives a lower limit for $g_i$ of ~ 6 Hz, since the implantation yield is unknown for this ESR centre. This compares to $g_i$ ~ 2 Hz for Gd implanted $Al_2O_3$ [35] and ~ 70 Hz for Er implanted $Y_2SiO_5$ crystal.[9] We observed no change in $\omega_r$ as $B_0$ was swept through the Er spin resonance, indicating the system is operating in the weak coupling regime. Our micro-resonator measurement represents the first reported coupling of a SC resonator to a RE ensemble implanted in Si. The strong coupling regime could be attained by optimization of the annealing recipe to produce only one ESR centre and by operating at higher centre frequencies.

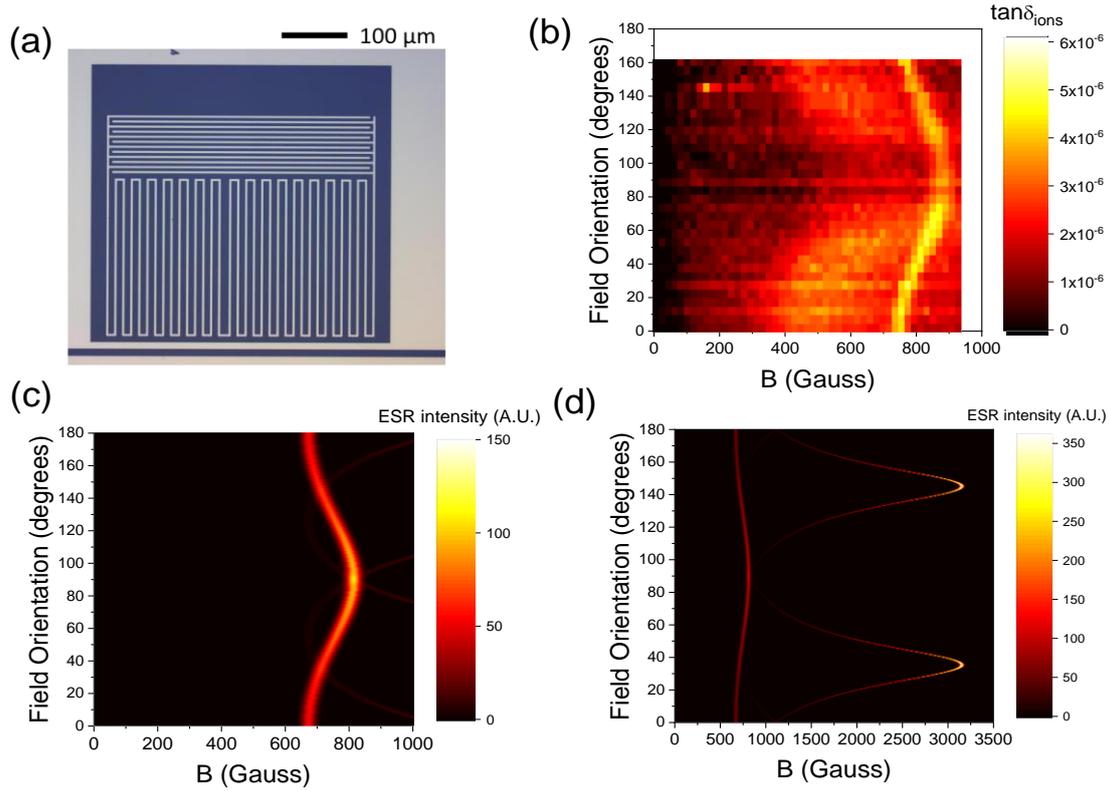

**Figure 6** a), Image of the superconducting micro-resonator that was coupled to sample 3 ($10^{17}$ Er cm$^{-3}$ and $10^{20}$ O cm$^{-3}$, annealing recipe *c*). b), Angular dependent micro-resonator ESR measurement of sample 3 at 20 mK. c), simulated angular dependent ESR spectrum using EASYSPIN numerical modelling for the trigonal OEr-2' centre identified by Carey *et al.*[14] with $g_\parallel$ = 0.69 and $g_\perp$ = 3.24. d), Simulated angular dependent ESR extended to higher $B_0$ to show the positions of the three expected ESR resonances with trigonal symmetry. The microwave frequency was 3.04 GHz for all micro-resonator measurements and simulations.

A promising scheme to develop a hybrid quantum computer is to link the long coherence time of silicon-based qubits with the ability to entangle superconducting qubits, which tend to have shorter coherence times. This requires the coupling of silicon and superconducting qubits which can be achieved by the exchange of a microwave photon.[36] These hybrid quantum circuits would exhibit long coherence times while allowing quantum state manipulation, and a quantum transducer could be developed using the ability of REs coupled to superconducting resonators to coherently convert optical photons to microwave photons using RE microwave Zeeman transitions as intermediaries. This would have applications in networking quantum signal processors, quantum key distribution and quantum metrology.

## 4. Conclusions

Er implanted Si is shown to be a promising platform for the development of QTs and is potentially highly scalable since it can utilise the silicon and ion implantation technology used in the IC industry. Er implanted Si can also exploit the atomic scale barrier to decoherence that is intrinsic to REs, and the recently developed ultra-low spin environment of isopically purified $^{28}$Si. Whereas the Er component itself is compatible with telecommunications wavelength photons and could be utilised for quantum communications schemes. We report the first coherence measurement of implanted Er in the form of a spin coherence time of ~10 μs at 5 K for $3\times10^{17}$ cm$^{-3}$ Er. The origin of this echo is an Er centre surrounded by six O atoms with monoclinic site symmetry. The spin echo decay profile had superimposed modulations due strong superhyperfine coupling with $^{29}$Si.

The temperature quenching of PL from a $10^{19}$ cm$^{-3}$ Er sample displayed previously unreported negative thermal quenching which implied the presence of an Er-O defect state below $^4I_{13/2}$ Er$^{3+}$ excited state, the unusually low PL lifetime of <1 μs is consistent with non-radiative decay enabled by this defect state. Fitting the temperature quenching of PL indicated that the $^4I_{15/2}$ Er$^{3+}$ ground state was buried in the valence band; this was also indicated by OMMR measurements. An ESR resonance under illumination at 1.5 μm that was three orders of magnitude stronger than unilluminated ESR resonances also implied the presence of an Er-O defect state below $^4I_{13/2}$ Er$^{3+}$ excited state to enable spin selective non-radiative decay to the Zeeman ground state. This mechanism is analogous to the optical spin polarisation mechanism of NV centres in diamond, and if correct could mean high temperature operation of Er qubits in Er implanted Si is possible.

We observed the first coupling between a superconducting resonator and Er implanted Si with $g_{col}$ = 1MHz and $g_i$ > 6 Hz, which provides a basis for future networks of hybrid quantum processors that exchange quantum information over the telecommunication network. Out of six known Er-related ESR centres, only one trigonal centre coupled to the SC resonator at 20 mK.

**Acknowledgements**

This work was supported by the UK EPSRC grants EP/R011885/1 and EP/H026622/1. We acknowledge the European Research Council for financial support under the FP7 for the award of the ERC Advanced Investigator Grant SILAMPS 226470. We would like to thank Prof. John Morton for helpful discussions.

**Data availability**

The datasets generated during the current study are available in the Mendeley Data repository

http://dx.doi.org/10.17632/s73x8nb8dn.1

**Supporting information**

Supporting Information is available from the Wiley Online Library or from the author.

**Competing financial interests:** The authors declare no competing financial interest

**Table of contents entry:**

Erbium implanted silicon is a promising quantum technology platform. This work shows the first coherence measurement of any implanted erbium to be ~10 µs at 5K. Three independent measurements all suggest that erbium implanted silicon has an energy level structure analogous to NV centres in diamond, which could lead to optical spin polarisation and high temperature operation of erbium qubits.

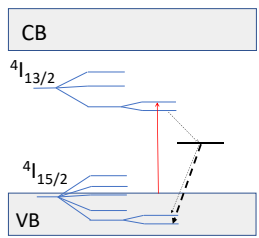